\documentclass[structabstract]{aa}  
\usepackage{graphicx}
\usepackage{txfonts}
\usepackage{subfigure}

\begin{document}
%
   \title{A $\delta$ Scuti star in the post-MS contraction phase: 44 Tauri}


   \author{P. Lenz\inst{1}
          \and
          A. A. Pamyatnykh\inst{1,2,3}
          \and
          T. Zdravkov\inst{2}
          \and
          M. Breger\inst{1}
          }

   \institute{Institut f\"ur Astronomie, University of Vienna,
              T\"urkenschanzstrasse 17, A-1180 Vienna, Austria\\
              \email{patrick.lenz@univie.ac.at}
              \and
              Copernicus Astronomical Centre, Bartycka 18, 00-716 Warsaw, Poland
              \and
              Institute of Astronomy, Russian Academy of Sciences, Pyatnitskaya Str 48, 109017 Moscow, Russia
             }

   \date{Received date; accepted date}

 
  \abstract
   {The evolutionary stage of the $\delta$ Scuti star 44 Tau has been unclear. Recent asteroseismic studies have claimed models on the main sequence, as well as in the expansion phase of the post-main sequence evolution. However, these models could not reproduce all of the observed frequencies, the mode instability range, and the fundamental stellar parameters simultaneously.
A recent photometric study has increased the number of detected independent modes in 44~Tau to 15, and a newly found gravity mode at 5.30~cd$^{-1}$ extends the observed frequency range.}
   { One of the possible evolutionary stages of 44~Tau has not yet been considered: the overall contraction phase after the main sequence. We computed asteroseismic models to examine whether models in this evolutionary stage provide a better fit of the observed frequency spectrum.}
   {We used Dziembowski's pulsation code to compute nonadiabatic frequencies of radial and nonradial modes. Observation of two radial modes and an avoided crossing of dipole modes put strong constraints on the models. A two-parametric overshooting routine is utilized to determine the efficiency of element mixing in the overshoot layer above the convective core.}
   {We find that pulsation models in the post-MS contraction phase successfully reproduce the observed frequency range, as well as the frequency values of all individual radial and nonradial modes. The theoretical frequencies of the mixed modes at 7.79~cd$^{-1}$ and 9.58~cd$^{-1}$ are in better agreement with the observations if efficient element mixing in a small overshoot layer is assumed.}
   {}


   \maketitle
%

\section{Introduction}

\begin{table*}[t!]
\caption{Observed frequencies and mode identifications.}
\label{table1}
\footnotesize
\begin{center}
\begin{tabular}{lcccccc}
\noalign{\smallskip}
\hline\hline
\noalign{\smallskip}
\multicolumn{2}{c}{Frequency} & A$_y$  & A$_{<v^1>}$  & $(\ell, m)$ & A$_v$/A$_y$  &  $\phi_v$-$\phi_y$   \\
 & [cd$^{-1}$] & [millimag]  & [km $s^{-1}$] & & &  [$^\circ$]    \\
\noalign{\smallskip}
\hline
\noalign{\smallskip}
$f_{1}$ &	6.8980  &  27.32 & 2.22	& (0,0) &  1.442 $\pm$ 0.003 &  +2.89 $\pm$ 0.13	\\
$f_{2}$ &	7.0060	&   9.38 & 0.46 & (1,1)	&  1.465 $\pm$ 0.011 &  -1.57 $\pm$ 0.43 \\
$f_{3}$ &	9.1174	&   8.39 & 0.44 & (1,1)	&  1.463 $\pm$ 0.011 &  -1.19 $\pm$ 0.45 \\
$f_{4}$ &	11.5196	&   9.93 & 0.73 & (1,0)	&  1.421 $\pm$ 0.009 &  -2.23 $\pm$ 0.36  \\
$f_{5}$ &	8.9606	&   9.31 & 1.03 & (0,0)	&  1.463 $\pm$ 0.012 &  +1.92 $\pm$ 0.45 \\
$f_{6}$ &	9.5611	&   8.50 & 0.30 & (1,-)	&  1.451 $\pm$ 0.012 &  -0.94 $\pm$ 0.51 \\
$f_{7}$ &	7.3031	&   4.74 & 0.70 & (2,0)	&  1.449 $\pm$ 0.025 &  -6.91 $\pm$ 0.90 \\
$f_{8}$ &	6.7955	&   2.58 & 0.28 & (2,0)	&  1.342 $\pm$ 0.041 &  -7.26 $\pm$ 2.01 \\
$f_{9}$ &	9.5828	&   1.84 & 0.12 & (2,-)	&  1.517 $\pm$ 0.054 &  -7.82 $\pm$ 2.02 \\
$f_{10}$ &	6.3390	&   1.69 & 0.21 & (2,-)	&  1.482 $\pm$ 0.064 &  -6.55 $\pm$ 2.24 \\
$f_{11}$ &	8.6391	&   1.78 & 0.32 & (-,0)	&  1.383 $\pm$ 0.054 &  -4.87 $\pm$ 2.19 \\
$f_{12}$ &	11.2947	&   1.03 & 0.11 & 	--	&       --                    &   --               \\	
$f_{13}$ &	12.6915	&   0.28 & --   &   --  &       --                    &  --                \\
$f_{14}$ &	5.3047	&   0.70 & --   &	--  &      --                     &  --                \\		
$f_{15}$ &	7.7897	&   0.28 & --   &	--  &      --                     &  --                \\		
\noalign{\smallskip}
\hline
\end{tabular}
\end{center}
\end{table*}

The $\delta$ Scuti stars are a class of pulsators exhibiting low-order acoustic and gravity modes driven by the opacity mechanism acting in the HeII ionization zone. In evolved $\delta$ Scuti stars mixed modes, i.e., modes with acoustic behaviour in the envelope and gravity behaviour in the interior, are common. Since such modes probe different layers inside a star than purely acoustic modes, the $\delta$ Scuti stars are excellent targets for asteroseismology.

44~Tau is a $\delta$~Scuti pulsator of spectral type F2 IV. Its
exceptionally slow intrinsic rotation of $V_{\rm rot}$ = 3 $\pm$ 2 km s$^{-1}$ (Zima et
al. 2007) is much lower than the average value for $\delta$ Scuti stars. For this group of pulsators, projected rotational velocities, $v \sin i$, around 100 km s$^{-1}$ and higher are not unusual. This makes 44~Tau an interesting and relatively simple target for asteroseismology, because the effects of rotation are weak and do not complicate mode identification and stellar modelling. 

In recent years several ground-based observing campaigns of the Delta Scuti Network (DSN) were organized to retrieve photometric and spectroscopic data of 44~Tau. The first comprehensive frequency analysis of its photometric light variations was published by Antoci et al. (\cite{antoci07}). This paper also summarizes all previous observations of the star. With two additional seasons of extensive Str\"omgren $vy$ photometry, Breger \& Lenz (\cite{breger08}) could increase the number of detected frequencies to 49, of which 15 are independent. 

Garrido et al. (\cite{garrido07}) have determined the spherical degree, $\ell$, for the 10 dominant modes by means of photometric amplitude ratios and phase differences. Their results were later confirmed by Lenz et al. (2008). In 2004 a spectroscopic campaign was organized simultaneously to the photometric observing run. The results of the line-profile variation analyses and the determination of photospheric element abundances are discussed in detail by Zima et al. (2007). For 8 modes, the azimuthal order, $m$, could be determined. Moreover, no indications of a global magnetic field and no anomalous element abundances were found.

The spectroscopically determined $\log g$ value of 3.6 $\pm$ 0.1 does not allow for unambiguous determination of the evolutionary status of 44~Tau. In recent studies, both main sequence (hereafter MS) and post-MS models in the expansion phase have been considered (Garrido et al. 2007, Lenz et al. 2008). The results by Lenz et al. (\cite{lenz08}) showed that these models fail to reproduce \emph{all} observable parameters satisfactorily (individual frequencies, observed frequency range, fundamental stellar parameters).

The effective temperature and luminosity of MS models is significantly lower than the values derived from photometry and Hipparcos data. Moreover, the positions of predicted $\ell$=2 modes do not match the observed frequencies (see Fig. 8 in Lenz et al. 2008). 

Evolved models in the hydrogen-shell burning stage reproduce the fundamental stellar parameters well, but theory predicts many more frequencies than actually observed. While partial mode trapping may explain the position of the $\ell$=1 modes, this mechanism should not be very effective for the observed $\ell$=2 modes. Moreover, the frequency range of unstable modes is shifted to higher frequencies with respect to MS models. Therefore, the mode at 6.34~cd$^{-1}$ and the newly found mode at 5.30~cd$^{-1}$ are predicted to be stable by models in the post-MS expansion phase.

We reexamined asteroseismic models of 44~Tau based on the latest frequency
solution of Breger \& Lenz (\cite{breger08}) and investigated a hitherto neglected possibility: a model in the overall contraction phase after the main sequence.  

During the main sequence evolution, a $\delta$ Scuti star transforms hydrogen into helium in its convective core. The amount of hydrogen in the core gets very low towards the end of this phase. The star increases its central temperature by an overall contraction to keep the energy production from hydrogen burning efficient. At the end of this overall contraction stage, hydrogen in the core is fully depleted and the convective core disappears\footnote{Of course, convective overshooting also disappears, but we keep the term `model with overshooting' or `the overshooting case' for later evolutionary stages to distinguish this case from models constructed without convective overshooting.}. A nuclear hydrogen-burning shell is established outside the hydrogen-exhausted core.
The stellar envelope expands again, which leads to a decrease of the effective temperature.

Finding a $\delta$ Scuti star in the contraction phase after the TAMS is not unlikely. If we consider the same range of effective temperatures, the evolution during the overall contraction phase is approximately 10 times faster compared to the MS evolution. In the post-MS expansion phase, the evolution is approximately 16 times faster than on the MS.  The asteroseismic inferences from post-MS contraction models are presented in this paper.

The paper is organized as follows. In Sect. 2 we briefly summarize the observed parameters of 44~Tau. In Sect. 3 we discuss the computation of our seismic models. The diagnostic power of mixed modes is examined in Sect. 4. Finally, in Sect. 5 we present predictions of post-MS contraction models.


\section{Observed frequencies and mode identification}

All observed independent frequencies of 44~Tau are given in Table~\ref{table1}, along with their mode identifications, if available. The given A$_y$ amplitudes are  mean Str\"{o}mgren $y$ amplitudes derived from the five observing seasons between 2000 and 2006. The amplitudes for the first moments are taken from Zima et al. (\cite{zima2007}). The spherical degrees of the modes were determined by Garrido et al. (2007) and Lenz et al. (2008). We reinvestigated these mode identifications with the new full frequency solution of Breger \& Lenz (\cite{breger08}) based on photometric amplitude ratios and phase differences and using the method by Daszy\'nska-Daszkiewicz et al. (\cite{jdd2003}, \cite{jdd2005}). The previous results were confirmed. The new mean amplitude ratios and phase differences between the Str\"omgren $v$ and $y$ bands are listed in Table~\ref{table1}.  
These average values were computed from the annual solutions of the five observing seasons and weighted by the annual signal-to-noise ratio of each mode. Taking the observed amplitude modulation in 44~Tau into account, this weighting scheme is more adequate than the one used in Lenz et al. (\cite{lenz08}). For some modes, in particular for $f_9$, the uncertainties in $A_v/A_y$ and $\phi_v-\phi_y$ could be reduced significantly.

The spherical degree can be uniquely determined for the 10 dominant modes. The observed mean phase difference of $f_{11}$ puts this mode midway between the theoretical values of $\ell=1$ and $2$ modes (see Fig.~5 in Lenz et al. \cite{lenz08}), which prevents an unambiguous identification. The uncertainties in the phase differences of the low-amplitude modes $f_{12}$--$f_{15}$ are too large to derive reliable results.

The mode positions in the diagnostic diagrams for mode identification depend on the efficiency of convection. By fitting the predicted mode positions to the observed values, Lenz et al. (2008) determined the mixing-length parameter, $\alpha_{\rm MLT}$, to be $\lesssim 0.2$ in 44~Tau. This is also confirmed by the new data. Consequently, convection in the envelope of 44~Tau is not very efficient.

The azimuthal order of eight modes was derived by Zima et al. (\cite{zima2007}) using the moment method. Most modes in 44~Tau are axisymmetric with exception of the two prograde modes $f_2$ and $f_3$. No rotational splitting has been observed. The close pair $f_6$ and $f_9$ at 9.56 and 9.58 cd$^{-1}$ does not represent a part of a rotationally split multiplet, because the determined spherical degrees differ. The inclination angle of 44~Tau could be constrained to 60 $\pm$ 25$^{\rm o}$.

The effective temperature of 44~Tau is $T_{\rm eff}$~=~6900~$\pm$~100~K, $\log L$~=~1.305~$\pm$~0.065, and $\log g$~=~3.6~$\pm$~0.1. Due to the use of a more recent estimate for the bolometric correction the given mean luminosity is lower than the value in Lenz et al. (2008). 

\begin{table*}[t!]
\caption{Parameters of post-MS contraction models with a good fit of the 15 frequencies observed in 44~Tau.}
\label{tab:bestmodel}
\footnotesize
\begin{center}
\begin{tabular}{cccccccccccc}
\noalign{\smallskip}
\hline\hline
\noalign{\smallskip}
Model & Opacity & Element mixture & X & Z & M [M$_\odot$] & $w$ & $\alpha_{\rm ov}$ & $\log T_{\rm eff}$ & $\log L$ & $\lg g_{\rm eff}$ & $\chi^2$\\
\noalign{\smallskip}
\hline
\noalign{\smallskip}
1 & OPAL  & GN93 & 0.70 & 0.02 & 1.860 & 8.0 & 0.178 & 3.8225 & 1.2777 & 3.6707 &  0.0128\\
2 & OPAL  & GN93 & 0.72 & 0.02 & 1.920 & 2.0 & 0.268 & 3.8200 & 1.2773 & 3.6753 &  0.0383\\
3 & OPAL  & GN93 & 0.72 & 0.02 & 1.922 & 8.0 & 0.155 & 3.8215 & 1.2839 & 3.6749 &  0.0081\\
4 & OPAL  & GN93 & 0.75 & 0.02 & 2.021 & --  & 0.195 & 3.8220 & 1.3016 & 3.6823 &  0.0077\\
5 & OPAL  & A04 & 0.74 & 0.012 & 1.783 & 8.0 & 0.219 & 3.8077 & 1.2091 & 3.6621 &  0.0623\\
6 & OPAL  & A04 & 0.80 & 0.012 & 1.958 & 8.0 & 0.164 & 3.8060 & 1.2292 & 3.6760 &  0.0145\\
\noalign{\smallskip}
7 & OP    & GN93 & 0.70 & 0.02 & 1.678 & 8.0 & 0.212 & 3.7784 & 1.0711 & 3.6561 &  0.0581\\
8 & OP    & GN93 & 0.75 & 0.02 & 1.800 & 8.0 & 0.166 & 3.7740 & 1.0745 & 3.6660 &  0.0619\\
\noalign{\smallskip}
9 & enh. OP & GN93 & 0.70 & 0.02 & 1.839 & 8.0 & 0.185 & 3.8190 & 1.2608 & 3.6688 &  0.1222\\
 10 &  enh. OP &  A04 &  0.74 &  0.012 &  1.812 &  8.0 &  0.213 &  3.8133 &  1.2361 &  3.6641 &  0.0360\\
 11 &  enh. OP &  A04 &  0.80 &  0.012 &  1.981 &  8.0 &  0.158 &  3.8098 &  1.2485 &  3.6770 &  0.0114\\
\noalign{\smallskip}
\hline
\end{tabular}
\end{center}
\end{table*}

\section{Seismic models in the post-MS contraction phase}

In this study we used the Warsaw-New Jersey evolutionary code, which is a descendant of the Paczy{\'n}ski (\cite{pacz1970}) code. In this code, envelope convection is described with the standard mixing-length theory of convection. 
The evolutionary models are computed assuming uniform rotation and conservation of global angular momentum during evolution from the ZAMS.

Aside from the traditional description of overshooting from the convective core, we also applied a new two-parametric description that allows for partial element mixing in the overshooting region (Dziembowski \& Pamyatnykh \cite{wad08}). It allows us to consider different profiles of the hydrogen abundance inside the partly mixed region just above the convective core.
One of the parameters, $\alpha_{ov}$, defines the overshooting distance above the convective core. The second parameter, $w$, adjusts the profile of the hydrogen abundance between the convective core and the upper boundary of the overshooting region. It therefore characterizes the efficiency of mixing in this region. The hydrogen mass fraction in a stellar layer, X, is defined as a function of fractional mass, $q$:
\begin{equation}
  X = X_c + (q - q_c)^w [a+b(q-q_c)].
\end{equation}
The coefficients $a$ and $b$ are determined from $w$ and $\alpha_{\rm ov}$. The subscript $c$ refers to the central values of the given quantities. An infinitely high value of $w$ corresponds to the traditional standard treatment for overshooting.

To compute the nonadiabatic radial and nonradial frequencies, we used the
most recent version of Dziembowski's pulsation code (\cite{wad77}). 
This code relies on the frozen convective flux approximation. The effects of rotation on the frequencies were treated up to second order following Dziembowski \& Goode (\cite{wad92}).

In this study we used two sources of the Rosseland mean opacities: the OPAL
opacity tables (Iglesias \& Rogers \cite{iglesias1996}) and the OP opacities (Seaton \cite{seaton2005}). At low temperatures, these tables are supplemented by the Ferguson \& Alexander (\cite{ferg05}) molecular and grain opacities.

The abundances of some photospheric elements in 44~Tau were determined by Zima
et al. (2007). Within the given uncertainties, the abundances of 44~Tau are close to the solar values by
Grevesse \& Noels (\cite{gn93}, hereafter GN93) and Asplund et
al. (\cite{asplund04}, \cite{asplund05}, hereafter A04). Since only abundances of 11 elements were measured, we did not utilize a special element mixture for 44~Tau, but examined pulsation models with both solar element mixtures.

\subsection{Results with OPAL opacities}

Two radial modes are observed in 44~Tau. As discussed in Lenz et al. (2008), this allows us to utilize Petersen diagrams (Petersen \& J{\o}rgensen 1972). In these diagrams, the period ratio between two radial modes is plotted against the period of the fundamental mode. The period ratio is sensitive not only to the adopted metallicity but also to rotation as stated by Su{\'a}rez et al. (\cite{suarez2006}). Since these quantities are determined well by observations, we can use Petersen diagrams to find the optimum mass of a model. 

Another important observation puts tight constraints on the models: the small frequency separation of the two $\ell=1$ modes at 9.11 and 9.56~cd$^{-1}$ indicates that these modes undergo an avoided crossing (Aizenman et al. 1977). Since both modes are of mixed character, they are also sensitive to the extent of overshooting from the convective core. 

To find the optimum fit for the dipole modes, a series of computations with different overshooting distances, $\alpha_{\rm ov}$, and different efficiencies of  partial element mixing in the overshoot layer ($w$=2.0 and 8.0) were made. Models that exhibited a good fit of all dipole modes also showed very good agreement for the quadrupole modes. For models obtained with X=0.70, the frequency separation between the dipole modes at 9.11 and 9.56~cd$^{-1}$ is predicted to be slightly larger than observed. An increase in the initial hydrogen mass fraction, X, helps to decrease the predicted frequency separation to the observed values and provides the best agreement for all observed 15 frequencies.
 
The best pulsation models for OPAL opacities and the GN93 and A04 element mixture are given in Table~\ref{tab:bestmodel}.
All models were computed with a mixing length parameter $\alpha_{\rm MLT}$=0.2. Model~4 was computed with the traditional description of overshooting. Starting with an initial rotation rate of 3.5~km~s$^{-1}$, the tabulated evolved models exhibit rotation rates close to 3~km~s$^{-1}$. The effects of rotation were also taken into account in the calculation of the pulsation frequencies.

\begin{figure}
   \centering
   \includegraphics[width=9cm, bb=10 15 290 218]{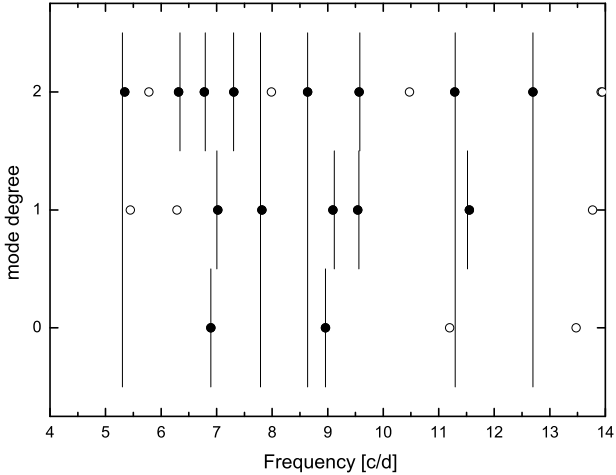}
      \caption{Comparison between observed frequencies (vertical lines) and
        predicted unstable frequencies (circles) for a post-MS contraction model (Model 4 in Table~\ref{tab:bestmodel}). If the spherical degree of a mode was not observationally determined, a full vertical line is shown. Predicted frequencies that match observed frequencies are marked as filled circles. The rotational splitting is smaller than the size of the symbols.
              }
         \label{fig:freq}
   \end{figure}

A comparison between predicted and observed frequencies for Model~4 in Table~\ref{tab:bestmodel} is shown in Fig.~\ref{fig:freq}. It can be seen that this pulsation model allows all observed modes to be explained solely with $\ell \leq 2$ modes. In some cases modes with $\ell$=3 or 4 are also close to observed values, but photometric cancellation effects clearly favour low degree modes.

To describe the goodness of the fit, we used a dimensionless merit function similar to the one used in Brassard et al. (\cite{brassard2001}):
\begin{equation}
  \chi^2 = \sum_{i=1}^{15} \left( \frac{\nu_{\rm obs}^i-\nu_{\rm model}^i}{\sigma^i} \right)^2
\end{equation}
where $\nu_{\rm obs}^i$ and $\nu_{\rm model}^i$ are the observed and corresponding theoretical frequency, respectively. The weighting factor, $\sigma^i$, is defined as the width of the observed frequency window of 5-13~cd$^{-1}$ divided by the number of theoretically predicted modes within this frequency window. Contrary to Brassard et al. (\cite{brassard2001}) we determine a weighting factor for each spherical degree separately to take the different mode densities into account. For the models shown in Table~\ref{tab:bestmodel}, $\sigma^i$ for $\ell=0$ modes amounts to $8/3$, for $\ell=1$ modes $8/7$ and for $\ell=2$ modes $8/11$~cd$^{-1}$. The results are shown in the last column of Table~\ref{tab:bestmodel}. The lowest $\chi^2$ value provides the best agreement between the model and the observations. If the azimuthal order of a mode was not uniquely determined, the theoretical $m=0$ mode was used to compute $\chi^2$. If the m-component closest to the observed frequency is used instead, the $\chi^2$ values are slightly lower but exhibit the same general trends.

\begin{figure*}
     \centering
	 \subfigure[Models constructed with the GN93 element mixture.]{
       \includegraphics[width=8.5cm, bb=10 15 275 225]{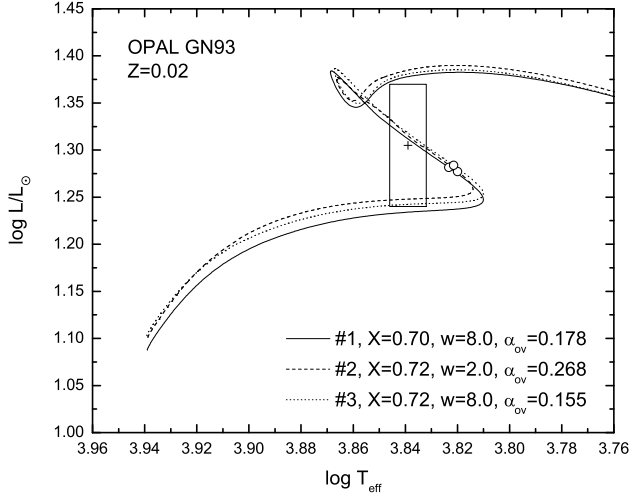}
	 }
	 \hspace{3mm}
	 \subfigure[Models constructed with the A04 element mixture.]{
       \includegraphics[width=8.5cm, bb=10 15 275 225]{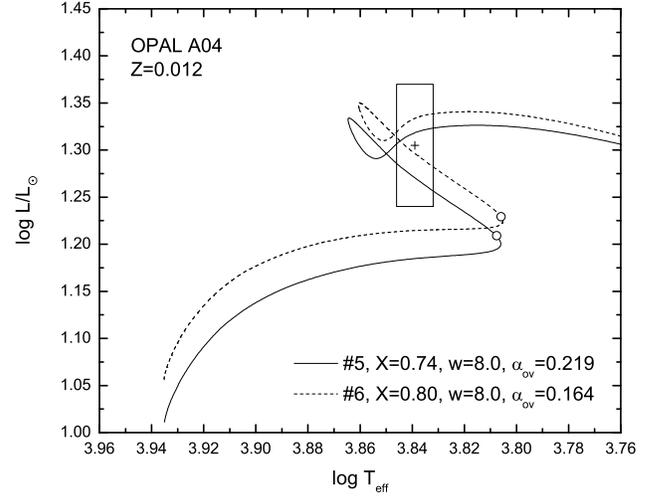}
	 }
     \caption{HR diagram with evolutionary tracks for models with a good fit of all 15 observed modes. The models were constructed with OPAL opacities and two different solar element mixtures: the GN93 element mixture (a) and the A04 element mixture (b).}
     \label{fig:hrdopal}
   \end{figure*}

The position of OPAL models in the HR diagram is shown in Fig.~\ref{fig:hrdopal} for models obtained with the GN93 and the A04 element mixture. The effective temperatures of these pulsation models are somewhat cooler than the values derived from photometric and spectroscopic measurements. 

The pulsation models obtained with the A04 mixture are closer to the TAMS. A 15-frequency fit with $w$=2.0 is not possible, because the required overshooting distance would be less than possible for models in the overall contraction phase. The main reason for the differences between pulsation models obtained with the GN93 and the A04 mixture is the slightly different opacity in the overshoot layer that affects the size of the convective core. The uncertainties in the element abundances (and opacities) in 44~Tau therefore lead to an uncertainty in the determination of the overshooting distance, $\alpha_{\rm ov}$.

\begin{figure}
   \centering
   \includegraphics[width=9cm, bb=10 15 290 218]{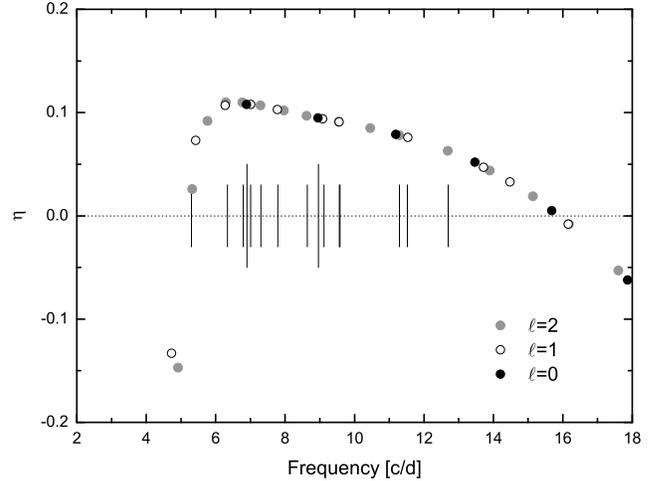}
      \caption{Instability parameter, $\eta$, for Model 3. Positive values indicate mode driving, modes with negative $\eta$ are damped.
              }
         \label{fig:eta}
   \end{figure}

The predicted frequency range of unstable modes agrees well with the observed frequency range as shown in Fig.~\ref{fig:eta} for one of the models that can be considered representative of all OPAL models. Additional unstable modes are predicted between 13 and 17 cd$^{-1}$. The mode instability at these high frequencies is sensitive to the efficiency of convection. If $\alpha_{\rm MLT}$ is decreased from 0.2 to lower values, the highest frequencies become stable. In our study we relied on the standard mixing-length theory of convection and the frozen flux approximation. Time-dependent convection models, such as those used in some  other recent studies about $\delta$ Scuti stars (e.g., Montalb{\'a}n \& Dupret \cite{mont2007}, Dupret et al. \cite{dupret2005}), would yield a more accurate determination of the high-frequency border of instability.

\subsection{Results with OP opacities}

\begin{figure}
     \centering
     \includegraphics[width=8.8cm, bb=9 7 150 173]{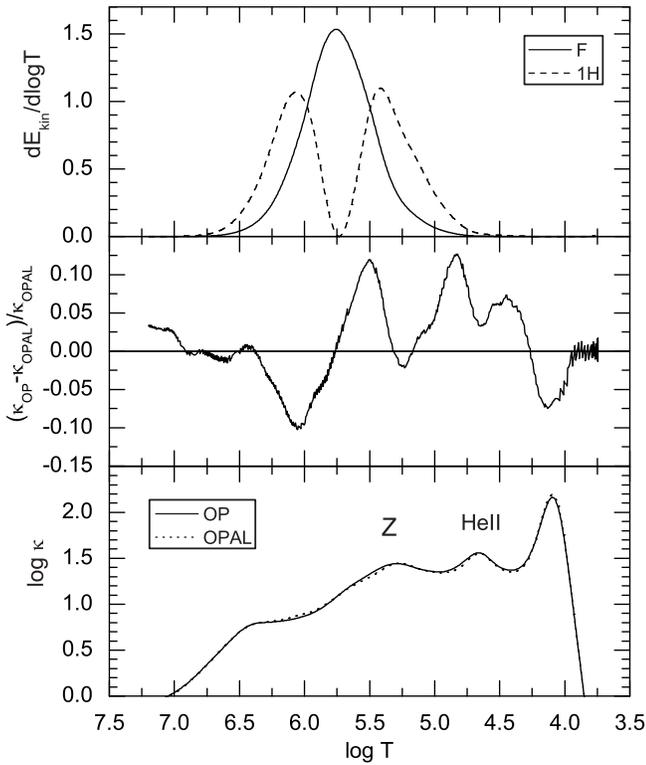}
     \caption{Radial fundamental and first overtone mode as probes for opacities. Upper panel: kinetic energy density inside the star. Middle panel: comparison between OPAL and OP opacities for the same model. Lower panel: absolute values of Rosseland mean opacities from OPAL and OP data. }
     \label{fig:opaccompar}
   \end{figure}

Lenz et al. (\cite{lenz07}) found that post-MS expansion models of 44~Tau constructed with OP opacities have  significantly lower temperature and luminosity values than observed. This problem persists for post-MS contraction models. Montalb\'{a}n \& Miglio (\cite{mont08}) explain this by differences of 10\% between OPAL and OP opacities at temperatures around $\log T$~=~6.05. This corresponds to the temperature region in which the period ratio is a sensitive probe as can be seen in the two uppermost panels in Fig.~\ref{fig:opaccompar}. 

The diagram shows the kinetic energy density of the radial fundamental and first overtone mode inside the star. Pulsation modes are sensitive to the conditions in temperature regions in which the kinetic energy density is high.
Because of the position of its node, the radial first overtone mode probes the temperature region between $\log T$~=~4.5 and 7.0 with different weights than the radial fundamental mode. 

The differences between OPAL and OP opacities are shown in the middle panel of Fig.~\ref{fig:opaccompar}. The lower panel shows the location of opacity bumps inside the star. We artificially enhanced the OP opacities by up to 15\% around $\log T$~=~6.0 and tested the impact on models in the contraction phase after the main sequence.

   \begin{figure}
     \centering
     \includegraphics[width=8.5cm, bb=10 15 275 225]{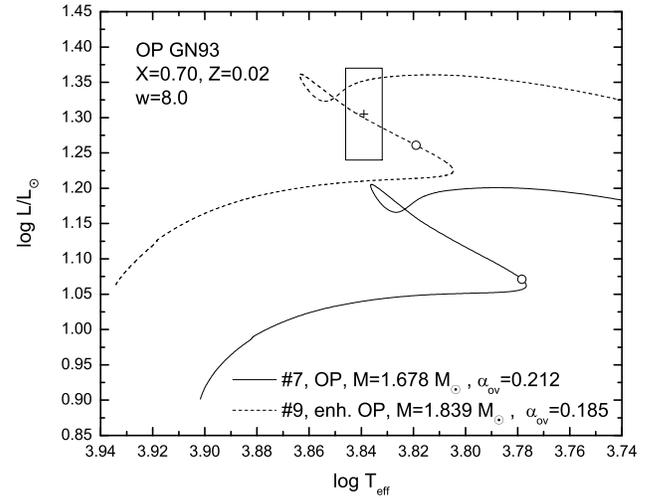}
     \caption{HR diagram with evolutionary tracks of two models constructed with enhanced and standard OP opacities. Both models provide a good 15-frequency fit.}
     \label{fig:hrdop}
   \end{figure}

The results are shown in Fig.~\ref{fig:hrdop}. The corresponding model parameters can be found in Table~\ref{tab:bestmodel}. Contrary to models constructed with standard OP opacities, models obtained with enhanced OP opacities are closer to the photometric error box in the HR diagram. This may indicate that the Rosseland mean opacities of the OP tables are currently underestimated in the temperature region around $\log T=$~6.0. 

For models constructed with standard OP opacities, a fit of all 15 frequencies is only possible with the GN93 element mixture. Adopting OP instead of OPAL opacities shifts the models that fit all 15 frequencies closer to the TAMS (e.g., compare the models in Fig.~\ref{fig:hrdopal}a with Model~7 in Fig.~\ref{fig:hrdop}). 
The models constructed with OPAL A04 are already situated at the TAMS (Fig.~\ref{fig:hrdopal}b). To fit the avoided crossing of $\ell=1$ modes and the $\ell=2$ modes in standard OP A04 models, $\alpha_{\rm ov}$ would have to be reduced more than possible for a model in the overall contraction phase. However, the use of enhanced OP opacities in combination with the A04 element mixture allows for a 15-frequency fit. The corresponding models are listed in Table~\ref{tab:bestmodel}.

\section{Mixed modes as probes of the stellar core}

The partial mixing processes at the convective core boundary are currently not fully understood. During the hydrogen core burning phase, we expect a region around the convective core that is partially mixed due to several mechanisms: (i) overshooting from the convective core, (ii) rotationally induced mixing (the core rotates faster than the envelope leading to additional mixing).

Unfortunately, until now no rotational splittings could be detected in 44~Tau. Therefore, we cannot derive the profile of differential rotation that would be necessary to disentangle the effects of rotation in element mixing.

An excellent probe of the size of the overshoot layer is the $g_1$ mode for $\ell$~$>$~0 as stated by Dziembowski \& Pamyatnykh (1991). This mode is partially trapped in the overshoot region, so its frequency has high diagnostic value. However, in 44~Tau models in the contraction phase after the TAMS, this mode already moved to higher frequencies outside the observed frequency range.

However, we observe many mixed modes in 44~Tau which are also sensitive to the conditions in the chemically inhomogeneous overshoot layer above the convective core. Table~\ref{tab:predictions} lists the fraction of oscillation kinetic energy confined in the g-mode cavity, $E_{\rm kin,g}/E_{\rm kin}$, for all predicted unstable modes in Model 3. Since the given values of $E_{\rm kin,g}/E_{\rm kin}$ are very similar for the different OPAL and OP models, they can be considered representative of all post-MS contraction models. Moreover, the differences between calculated and observed frequencies are given. If the azimuthal order of an observed mode is unknown, the frequency difference to the theoretical $m$=0 modes is given and the value is enclosed by brackets. The uncertainty in frequency due to unknown azimuthal order may be as high as 0.05 cd$^{-1}$ for $\ell=2$ modes.

\begin{table}[h!]
\caption{Theoretical frequencies and mode properties of unstable modes predicted by Model 3.}
\label{tab:predictions}
\footnotesize
\begin{center}
\begin{tabular}{lccccc}
\noalign{\smallskip}
\hline\hline
\noalign{\smallskip}
 \multicolumn{2}{c}{Mode} & $\nu_{\rm model}$ & $\nu_{\rm observed}$ & $\nu_{\rm observed}-\nu_{\rm model}$ & $E_{\rm kin,g}/E_{\rm kin}$ \\
 $\ell$ & ID & [cd$^{-1}$] & [cd$^{-1}$] & [cd$^{-1}$] &  \\
\noalign{\smallskip}
\hline
\noalign{\smallskip}
 0 & F  &  6.8980 & 6.8980 &  0.0000 & 0.0 \\
 0 & 1H &  8.9607 & 8.9606 & -0.0001 & 0.0 \\
 0 & 2H & 11.20   & --     &  --     & 0.0 \\
 0 & 3H & 13.48   & --     &  --     & 0.0 \\
\noalign{\smallskip}
 1 & g$_6$ &  5.44   &  --     & --               & 0.96\\
 1 & g$_5$ &  6.29   &  --     & --               & 0.92\\
 1 & p$_1$ &  7.0342 &  7.0060 & -0.0282          & 0.13 \\
 1 & g$_4$ &  7.7915 &  7.7897 & \emph{(-0.0018)} & 0.87 \\
 1 & p$_2$ &  9.1099 &  9.1174 & 0.0075           & 0.35 \\
 1 & g$_3$ &  9.5748 &  9.5611 & \emph{(-0.0137)} & 0.61 \\
 1 & p$_3$ & 11.5526 & 11.5196 & -0.0330          & 0.04 \\
 1 & p$_4$ & 13.74   &  --     & --               & 0.19 \\
 1 & g$_2$ & 14.51   &  --     & --               & 0.75 \\
\noalign{\smallskip}
 2 & g$_{10}$ & 5.3375 &  5.3047 & \emph{(-0.0328)}  & 0.96 \\
 2 & g$_9$ & 5.78      & --      & --               & 0.93 \\
 2 & g$_8$ & 6.3076    &  6.3390 & \emph{(0.0314)} & 0.83 \\
 2 & g$_7$ & 6.7795    &  6.7955 & 0.0160          & 0.68 \\
 2 & f     & 7.3023    &  7.3031 & 0.0008          & 0.67 \\
 2 & g$_6$ & 7.98      & --      & --               & 0.74 \\
 2 & p$_1$ & 8.6401    &  8.6391 & -0.0010           & 0.64 \\
 2 & g$_5$ & 9.5584    &  9.5828 & \emph{(0.0244)} & 0.62 \\
 2 & p$_2$ & 10.47     & --      & --               & 0.57 \\
 2 & g$_4$ & 11.3129   & 11.2947 & \emph{(-0.0182)}  & 0.60 \\
 2 & p$_3$ & 12.7043   & 12.6915 & \emph{(-0.0128)}  & 0.34 \\
 2 & g$_3$ & 13.92     & --      & --               & 0.04 \\
 2 & p$_4$ & 15.17     & --      & --               & 0.35 \\
\noalign{\smallskip}
\hline
\end{tabular}
\end{center}
\end{table}

The mixed $\ell$=1 mode at 7.79 cd$^{-1}$ has strong g-mode characteristics. Almost 87\% of its kinetic energy is confined in the g-mode cavity in the stellar interior. This also explains why the observed amplitude of this mode is significantly smaller than that of other dipole modes in 44~Tau. Consequently, this and other mixed modes are sensitive to the shape of the hydrogen profile in the partially mixed layer above the convective core.

   \begin{figure}
   \centering
   \includegraphics[width=9cm, bb=10 15 270 210]{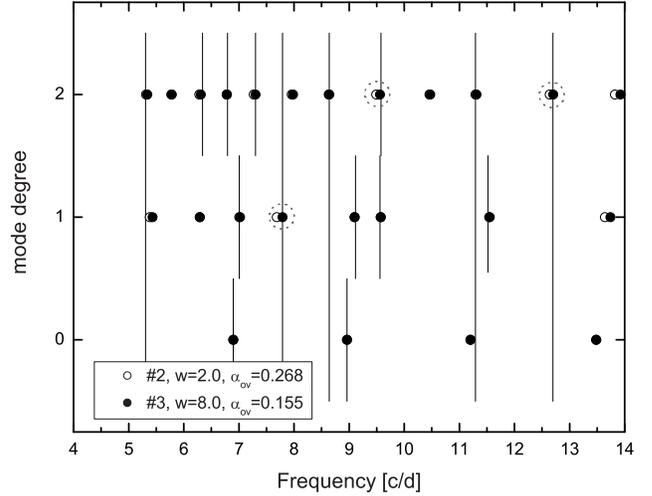}
      \caption{Comparison of the theoretical frequencies of two pulsation models with different hydrogen profile modelling, $w$=8.0 (filled circles) and $w$=2.0 (open circles). Modes that are most affected are marked with a dotted circle. The parameters of the models are given in Table~\ref{tab:bestmodel} (see Models 2 and 3).}
         \label{fig:w8w2compar}
   \end{figure}

We computed pulsation models using two different assumptions for element mixing in the overshoot layer: weak element mixing ($w$=2.0) and efficient element mixing ($w$=8.0). For comparisons we also calculated models with the standard description of overshooting, which assumes no element mixing in the overshooting region.

  In Fig.~\ref{fig:w8w2compar} we compare the predicted frequency spectra for Models 2 and 3 (obtained with $w$=2.0 and $w$=8.0, respectively). The corresponding overshooting distances were adjusted to values that provide the best fit of the observed frequencies. As can be seen, the frequencies of some modes are sensitive to the differences in the overshooting region. The frequency of the $g_{4}$ mode at 7.79~cd$^{-1}$ shifts most significantly. The frequency difference  $\Delta \nu$=0.1~cd$^{-1}$ is larger than the expected maximum frequency separation between components of a rotational splitting. Consequently, our results suggest preference for more efficient mixing in a small partially mixed region ($w$=8.0) over less efficient mixing in a larger partially mixed region ($w$=2.0).

   \begin{figure}
   \centering
   \includegraphics[width=9cm, bb=15 20 321 390]{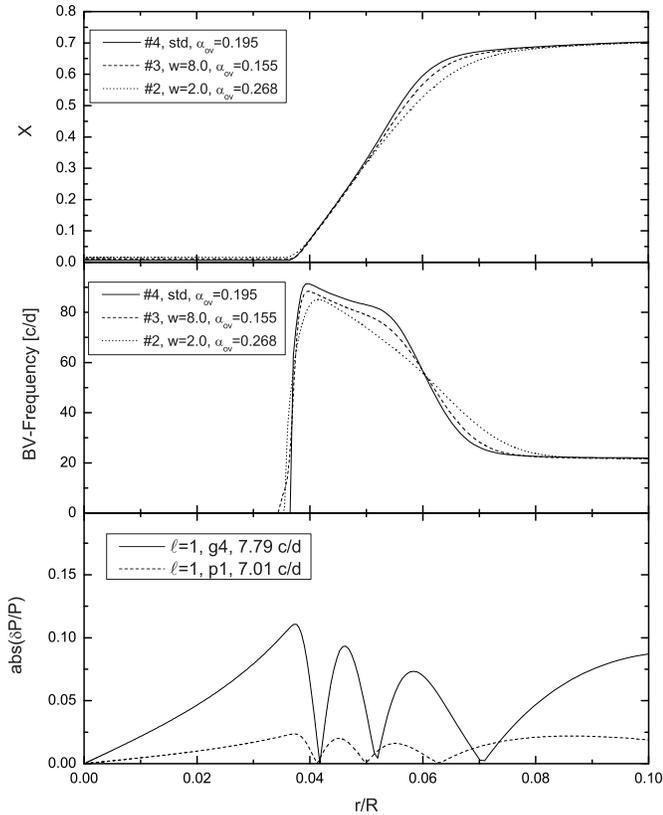}
      \caption{Impact of different efficiency of element mixing in the overshoot layer on the profile of the hydrogen abundance (upper panel) and on the Brunt-V\"ais\"al\"a frequency (middle panel). Lower panel: relative Lagrangian pressure perturbation for two mixed $\ell$=1 modes, the mainly acoustic mode at 7.01~cd$^{-1}$ and the mainly gravity-type mode at 7.79~cd$^{-1}$.
              }
         \label{fig:profile}
   \end{figure}

The corresponding profile of the hydrogen abundance for the discussed pulsation models is given in the upper panel of Fig.~\ref{fig:profile}. The different hydrogen profiles lead to a modification of the Brunt-V\"ais\"al\"a frequency, as shown in the panel in the middle of the diagram. In the lower panel the eigenfunction of the relative Lagrangian pressure variation is given for the mixed $\ell=1$ modes: $g_{4}$ at 7.79~cd$^{-1}$ and $p_{1}$ at 7.01~cd$^{-1}$. Due to its larger amplitude in the partially mixed region the gravity mode is very sensitive to the conditions in this region. The change in the Brunt-V\"ais\"al\"a frequency due to the different chemical profile slightly adjusts the size of the g-mode cavity and, therefore, also influences the values of the frequencies.

\section{Predictions of the post-MS contraction model}

Post-MS contraction models of 44~Tau predict a few additional dipole and quadrupole modes that may be found by future observations. After fitting the observed 15 modes, the frequencies of the remaining modes no longer strongly depend on the input parameters. However, small deviations may be found, especially for  g-modes that will provide additional constraints on mixing in the overshoot layer. The expected frequencies of hitherto not observed modes can be found in Table~\ref{tab:predictions}. The detection of these modes would help refining our models.

A reexamination of our photometric data indeed shows a prominent peak at the predicted position of the second radial overtone (11.198~cd$^{-1}$). With an SNR of 2.72, this peak is below the significance limit of 4.0 but may be confirmed with new additional data. A reexamination of radial velocity data did not show any reliable prominent peaks at the predicted position of modes. Many of the predicted modes that lack an observational counterpart are gravity modes for which we only expect low amplitudes. Therefore, more accurate data are needed to extract these frequencies.

Moreover, the model predicts the spherical degree of some of the observed modes: $\ell=1$ for 7.79~cd$^{-1}$ and $\ell=2$ for 5.30, 8.64, 11.30, 12.69~cd$^{-1}$. These predictions also need to be confirmed by high-resolution spectroscopic data.

The computed evolutionary relative period changes, (1/P)dP/dt, are approximately $1\cdot10^{-8}$ yr$^{-1}$. Such changes are too small to be measured with the 5 year time-base of our data.

\section{Conclusions}

We constructed seismic models of the $\delta$ Scuti star 44~Tau in the overall contraction phase after the main sequence. The two observed radial modes and an avoided crossing of a pair of dipole modes put strong constraints on pulsation models. Unlike models on the main sequence or in the post-MS expansion phase, these models successfully reproduce all 15 independent frequencies observed in 44~Tau and predict all detected modes unstable. This makes 44~Tau an example for determining the evolutionary stage by means of asteroseismology.

The observed frequency spectrum (obtained from ground-based data) can be solely explained with modes with spherical degrees of 0, 1 and 2. This result is what we expect for the given accuracy of the data.

The frequency spectrum of 44~Tau consists mainly of modes of mixed character, i.e., modes with acoustic behaviour in the envelope and gravity behaviour in the interior. These modes allow examination of the efficiency of partial element mixing in the overshoot layer around the stellar core. We find that the theoretical frequencies of several modes (e.g. the dipole $g_4$ mode at 7.79~cd$^{-1}$ and the quadrupole mode $g_5$ at 9.58~cd$^{-1}$) are in better agreement with efficient mixing in a thin overshoot layer than with less efficient mixing in an extended overshoot layer.

The main source of uncertainty in the asteroseismic modelling of 44~Tau are the stellar opacities. Determination of the size of the overshooting region above the convective core, therefore, depends on the source of opacity data and, to a smaller extent, on the choice of the element mixture.
Pulsation models in the post-MS contraction phase constructed with standard OP data are more than two standard deviations outside the photometric error box in the HR diagram. Following a suggestion by Montalb{\'a}n \& Miglio (\cite{mont08}), we computed a model with artificially enhanced OP opacities around the temperature region $\log T$~=~6.0. Such a model indeed agrees better with the fundamental parameters of 44~Tau derived from photometry and Hipparcos data.

The pulsation models presented in this paper not only provide an excellent fit of all observed modes but also allow several predictions, which may be confirmed with additional observational data. A few of the theoretically predicted modes are not observed. The question arises whether the amplitudes of these modes are below the noise threshold of the current data or whether they are even excited at all. Moreover, we predict the spherical degree of already observed modes for which our mode identification techniques did not succeed because the uncertainties in amplitudes and phases are too large. 
The excellent fit of the individual frequencies is very promising, and we are confident that additional data will lead to even more asteroseismic inferences about this star.

\begin{acknowledgements}
We thank Wolfgang Zima for searching for our predicted frequencies in the radial velocity data of 44~Tau.
 This investigation has been supported by the Austrian Fonds zur F\"orderung der wissenschaftlichen Forschung. AAP and TZ acknowledge partial financial support from the Polish MNiSW grant No. N N203 379636.
\end{acknowledgements}

\end{document}